# Operator-oriented programming: a new paradigm for implementing window interfaces and parallel algorithms

P.S. Ruzankin
Novosibirsk State University, Sobolev Institute of Mathematics

**Abstract**
We present a new programming paradigm which can be useful, in particular, for implementing window interfaces and parallel algorithms. This paradigm allows a user to define operators which can contain nested operators. The new paradigm is called operator-oriented. One of the goals of this paradigm is to escape the complexity of objects definitions inherent in many object-oriented languages and to move to transparent algorithms definitions.

## *Introduction*

The main goal of this paper is to suggest a new paradigm, which we call operator-oriented programming and which can be useful, in particular, for implementing window interfaces and parallel algorithms. This paradigm uses a more complex operator definition syntax, than the contemporary programming languages have. This allows defining operators with nested operators and operator inheritance. This complexity, however, leads to clarity of the structure of a program and simplicity of parallel constructs.

The "programming paradigm" term, though having no commonly accepted meaning, usually means a set of fundamental approaches to programming, or, as Wikipedia says, "a fundamental style of computer programming" [1]. Nevertheless there are many programming styles that are commonly called programming paradigms. One of them is object oriented programming [2], which proved to be very useful for coding graphical user interfaces. However this paradigm is not free from drawbacks. One of them is performance degradation compared to traditional functional programming (e.g. see [3]). Another disadvantage is that the object oriented programs appear to be very long and complex and very hard to read.

One of the goals of the new operator oriented paradigm is to avoid the enormous complexity of objects definitions inherent in many contemporary object-oriented languages (such as C++) and to move to transparent algorithms definitions. Though the both paradigms: object-oriented and operator-oriented can be implemented in one language, an operator-oriented language does not need complex object-oriented techniques.

## *1. Definition of an operator*

In examples below we will use a new programming language, which is neither yet implemented nor named. We use it to elucidate how the new paradigm can be implemented.

Let us show how the source code for a simple dialog window will look like:

```
program hello_world;
   begin dialog_window "Title";
      dialog_message "Hello, world!";
      dialog_ok_button;
   end dialog_window;
end program hello_world;
```

It is clear that this code displays a small window with the message "Hello, world!" and the "Ok" button. But what in general the operator `dialog_window` must do?

First, the operator looks through the nested operators, each of which is responsible for a part of the window, and asks for the minimal size sufficient for each of these parts. Then the operator `dialog_window` decides how many pixels it wants to give to each of these parts and asks each of the nested operators to paint the corresponding part with the corresponding size. Of course, this is not sufficient for a real implementation of a dialog window. We intentionally omit the operations needed to deal with keyboard and mouse events, repainting the window etc. We focus only on formatting issues to make our exposition simple.

The operator-oriented programming allows a user to define such complex operators. The definition of the operator `dialog_window` must look like the following:

```
operator dialog_window(title);
   string title;
   method execute;
      integer x_size,y_size,x,y,x_position,y_position;
      x_size:=0;
      y_size:=0;
      begin by_nested_operators;
         this_operator.get_min_size(x,y);
         x_size:=max(x_size,x);
         y_size:=y_size+y;
      end by_nested_operators;
      paint_dialog_window(title,x_size,y_size);
      x_position:=0;
      y_position:=0;
      begin by_nested_operators;
         this_operator.get_min_size(x,y);
         this_operator.paint_the_part(x_position,y_position,x_size,y);
         y_position:=y_position+y;
      end by_nested_operators;
   end method execute;
end operator dialog_window;
```

Here `string` and `integer` are declarations of variables of the corresponding types, `by_nested_operators` is the loop by nested operators, `this_operator` means that the current nested operator is referred to.

In the above example one of the key features of the new paradigm is introduced: *the loop by nested operators*. In the example this loop is used, first, to acquire the minimal sizes for the parts of the window and, second, to paint these parts.

We see that an operator contains *methods* which are procedures similar to methods in object-oriented programming. The operators which are responsible for the parts of the window must have at least two methods: `get_min_size` and `paint_the_part`.

In contrast to ordinary loops, each step of the loop by nested operators must be translated (into machine code or intermediary code) at compilation time. This, however, allows employing in such loops operators which have methods with given names and given parameters, not necessarily the operators which inherit one given operator. In object-oriented programming for an object to be used for a specific purpose it is necessary that the corresponding class inherit a given class. This is a key difference between the operator-oriented paradigm and the object-oriented paradigm: *outer similarity* of operators for operator-oriented paradigm versus *inner similarity* of objects for object-oriented paradigm.

There is the only method `execute` needed for the `dialog_window` operator. When a compiler comes across this operator in the source code of a procedure, the compiler merely puts into the resulting code a call for this method.

Let us conclude this section with one more example to illustrate how such technique makes formatting a window easy. Let us add “Cancel” button to the hello-world dialog.

```
begin dialog_window “Title”;
   dialog_message “Hello, world!”;
```

```
   begin window_part_row;
      dialog_cancel_button;
      dialog_ok_button;
   end window_part_row;
end dialog_window;
```

Here the `window_part_row` operator paints the nested elements in a row. (We assumed that `dialog_window` operator arranges nested elements in a column.)

## *2. Application to parallel programming*

Let us now define an operator which executes in parallel all its nested operators and waits until all these operators complete. The definition will be like the following:

```
operator parallel_execute;
   method execute;
      semaphore s;
      s:=num_nested_operators;
      begin by_nested_operators;
         begin new_thread(s);
            this_operator.execute;
         end new_thread;
      end by_nested_operators;
      wait_zero_semaphore s;
   end method execute;
end operator parallel_execute;
```

Here the constant `num_nested_operators` equals the number of nested operators; `new_thread` operator creates a new thread; the semaphore `s` is a system integer variable for inter-thread communication, which here decrements by 1 when the execution of each `new_thread`'s nested operator completes; and `wait_zero_semaphore` operator waits until the corresponding semaphore becomes zero.

The usage of the above defined operator is obvious:

```
begin parallel_execute;
   a_time_consuming_operator1;
   a_time_consuming_operator2;
   a_time_consuming_operator3;
end parallel_execute;
```

A similar construction can certainly be implemented by means of traditional functional programming, or traditional languages can be extended to include operators for parallel execution. (In fact, both approaches are widely used.) However the key advantage of the operator-oriented programming is the flexibility of implementing operators for parallelism. The above defined operator can be redefined *by a user* in a large variety of ways to adapt implementation to different execution environments. This allows effective compilation of a parallel algorithm for an individual execution environment with no need to rewrite the algorithm. For instance, if we have only one one-threaded processor and do not need to run the subtasks simultaneously, it is faster to execute the subtasks sequentially. For this we rewrite the above definition:

```
operator parallel_execute;
   method execute;
      begin by_nested_operators;
         this_operator.execute;
      end by_nested_operators;
   end method execute;
end operator parallel_execute;
```

Now the same operator `parallel_execute` executes its subtasks sequentially.

## 3. Inheritance of operators

Like the object-oriented programming assumes that objects can be inherited, the operator-oriented programming is to allow inheritance for operators. Such inheritance can easily be implemented both at syntax level and at compiler level: an operator adopts from another operator all the variables and methods, each of which can be redefined, and some other variables and methods can be added.

However, in contrast to object-oriented programming, the inheritance of operators is not a key feature of operator-oriented programming and serves for convenience only. This is because of the aforementioned outer similarity used in operator-oriented paradigm. As it was seen in Section 1, to be a proper nested operator, an operator needs only to contain methods with given names and arguments, no inheritance from a particular operator being needed.

## 4. Namespaces

Let us consider one more example.

```
integer answer;
begin dialog_window "Title";
   dialog_message "Please answer";
   begin window_part_row;
      begin dialog_button "Yes";
         answer:=1;
         close_dialog;
      end dialog_button;
      begin dialog_button "No";
         answer:=0;
         close_dialog;
      end dialog_button;
   begin window_part_row;
end dialog_window;
```

It is clear that this code shows the window with two buttons "Yes" and "No". When a user presses a button, the dialog window is closed with the corresponding result code assigned to the variable `answer`.

One of the problems of implementation of the new language is to correctly define the namespace context. The assignment operator `answer:=1` needs the variable `answer` defined outside the `dialog_window` operator. And the `close_dialog` operator needs to know which dialog to close.

The simplest way to pass the needed context to the `close_dialog` operator is to pass it via parameters to the method `execute` of this operator. But this way is undesirable because this operator is used together with operators that have no parameters for the `execute` method.

Thus it would be more properly to allow an operator to share its variables and methods, i.e. to allow an operator to create a context for nested operators.

## 5. Disadvantages of the paradigm

The main disadvantage of the operator-oriented programming is that probably none of the existing compilers can be easily adapted to support this paradigm. A new compiler must be developed to implement the paradigm.

The other disadvantage is the direct consequence of the outer similarity of operators used. The same effects must arise as when performing compile-time inlining of procedures in functional programming. The resulting machine code may be greater in size than the corresponding code obtained from an object-oriented program. But the operator-oriented program may run faster than its object-oriented analog.

Authors' address: P. S. Ruzankin,
Novosibirsk State University, Sobolev Institute of Mathematics,
Pr. Ak. Koptyuga, 4,
Novosibirsk, 630090,
Russia
Email: ruzankin@math.nsc.ru